# Let There be Light: How Satisfiability Emerges from Periodic Shadows[1]

**Douglas Powell**

**Abstract**. **The 2-SAT and 3-SAT problems are keystones of computational complexity research, foundational for thousands of related problems. Here, I do not propose an equivalent complexity class for the two domains. Rather, I show that 2-SAT and 3-SAT share the same structure, a *k*-based search space. With $w = 2^k$, each of m clauses has *w k*-ary sub-assignments, giving $m(m-1)/2$ clause pairs and $w^2m(m-1)/2$ sub-assignment pairs, each compatible or incompatible according to the Boolean relations of its members. A candidate assignment is an ordered *m*-ary set of sub-assignments, one per clause, and is satisfying iff it contains no incompatible sub-assignment pair. Under a lexicographical ordering of the $w^m$ candidate assignments over the integer interval $[0, w^m - 1]$, each sub-assignment pair appears in a doubly-periodic set of exactly $w^{(m-2)}$ candidate assignments, described exactly by a doubly-periodic interval. The locations of all unsatisfying candidate assignments are therefore described by no more than $w^2m(m-1)/2$ doubly-periodic intervals, at $O(w^2m(m-1)/2)$ cost, although determining whether those intervals cover the search space appears non-trivial. The intervals may be cast as a system of light and shadows in which a candidate assignment is satisfying if and only if it is unshadowed. Thus viewed, the number of unsatisfying candidate assignments may be computed by inclusion-exclusion from $O(m^3)$ coefficients, given an oracle for those coefficients whose cost is not bounded here, and equivalently by a recursion on shadows. I propose investigating whether these intervals, rendered as superposed periodic signals, allow machine learning to discern satisfiability, and whether models trained on 2-SAT transfer to 3-SAT.**

## 1 Definition of 3-SAT

As characteristically described, a 3-SAT problem consists of a set of *n* Boolean ***variables*** $X = (X_0 \ldots X_{n-1})$, and the conjunction of *m* ***clauses***, $C = (c_0, \ldots, c_{m-1})$. (I rely on [CMS97] and [GP97] for the definitions in this section.) A ***literal*** $x_i$ or $\neg x_i$ is the true or false instantiation of variable $X_i$. Each clause is the disjunction of an ordered set of exactly three unique, non-tautological literals. A clause $(x_a \vee x_b \vee x_c)$ may be treated as the ordered set, $(x_a, x_b, x_c)$. Throughout this paper, let *F* be a 3-SAT problem in *m* clauses. An ***assignment*** is an ordered set of *n* literals, one literal for each variable of *X*. Assignment *A* ***satisfies*** clause $c_i$ if and only if:

1. *A* is non-tautological, and
2. $(A \cap c_i) \neq \emptyset$. (1)

Assignment *A* ***satisfies*** 3-SAT problem *F* iff *A* satisfies all *m* clauses of *F*, which requires that

1. $|A| = n$ (*all n variables are assigned*);
2. $\forall x_i \in X\,(\mathit{if}\,(\neg x_i \in A) \rightarrow (x_i \notin A))$ (*all literals are consistent* (non-tautological)); and
3. $\forall c \in C\,((A \cap c) \neq \emptyset)$ (*each clause is satisfied*). (2)

Determining whether a given 3-SAT problem is satisfiable may require testing all $2^n$ assignments, a generally intractable task. (See [MRPGTR16], [PVNP16], [FL09].)

## 2 Definitions for the Sub-Assignment Expansion

Here I show an expansion of 3-SAT in which the search space has a periodic structure such that the locations of all unsatisfying candidate assignments are known *a priori*. That conclusion relies on the following properties and definitions. Let **F** symbolize false and **T** symbolize true. The set of three Boolean literals of every 3-SAT clause is one of $2^3$ possible 3-ary forms:

$$((\mathbf{F}, \mathbf{F}, \mathbf{F}), (\mathbf{F}, \mathbf{F}, \mathbf{T}), (\mathbf{F}, \mathbf{T}, \mathbf{F}), (\mathbf{F}, \mathbf{T}, \mathbf{T}), (\mathbf{T}, \mathbf{F}, \mathbf{F}), (\mathbf{T}, \mathbf{F}, \mathbf{T}), (\mathbf{T}, \mathbf{T}, \mathbf{F}), (\mathbf{T}, \mathbf{T}, \mathbf{T})). \tag{2.1}$$

[1] This paper is a revised version of a previous manuscript by the author.

For brevity, the eight ***clause forms*** of (2.1) are symbolized: (**FFF, FFT, FTF, FTT, TFF, TFT, TTF, TTT**), indexed from 0…7. Thus, $c_i = (\neg x_0 \vee \neg x_3 \vee \neg x_7)$ is clause form $f = 0$, symbolized as **FFF**; but $c_i = (x_0 \vee \neg x_3 \vee x_7)$ is clause form $f = 5$, which is **TFT**. Likewise, every three-literal assignment to the three variables of each clause is one of those eight forms, symbolized and indexed the same. With $0 \leq s < 8$, the $s$-th three-literal assignment to any clause $c_i$ is called a ***sub-assignment***, denoted ***i.s*** for its clause subscript $i$ and sub-assignment form $s$. E.g., clause $c_i = (x_0 \vee \neg x_3 \vee x_7)$ of form $f = 5$, which is **TFT,** has sub-assignment $i.3 = (\neg x_0, x_3, x_7)$, which is **FTT**. Each of $m$ clauses has eight sub-assignments. Thus, every 3-SAT problem in $m$ clauses has $8m$ sub-assignments. Because no sub-assignment is tautological, sub-assignment *i.s* ***satisfies*** clause $c_i$ if and only if $(i.s \cap c_i) \neq \emptyset$. (See (1).) For clause $c_i$ of form $0 \leq f < 8$, there is exactly one unsatisfying ***anticlause*** sub-assignment, $i.(7 - f)$, (or $i.(3 - f)$ for 2-SAT). Exhaustive inspection of the foregoing sets shows that for any clause $c_i$ of form $0 \leq f < 8$: **(1)** every non-anticlause sub-assignment $i.s \in (i.0, ..., i.7) \setminus i.(7 - f)$ ***satisfies*** clause $c_i$; but **(2)** no anticlause sub-assignment $i.(7 - f)$ satisfies its clause, $c_i$. Sub-assignments $i.s \neq i.t \in (i.0, ..., i.7)$ are ***sibling*s**. By inspection, the union of literals of any pair of siblings is tautological under disjunction and contradictory (inconsistent) under conjunction.

Every 3-SAT problem in $m$ clauses has $m(m - 1)/2$ ordered ***clause pairs***, $(c_i, c_j)$, with clause subscripts $0 \leq i < j < m$. For clause subscripts $0 \leq i < j < m$, and sub-assignment forms $0 \leq s, t < 8$, $(i.s, j.t)$ is a ***sub-assignment pair***, i.e., an ordered pair of sub-assignments drawn from two different clauses. A sub-assignment pair is interpreted as the union of literals of its constituent sub-assignments. Let $ap = (a.b, c.d)$ and $bp = (e.f, g.h)$ be sub-assignment pairs. Sub-assignment pairs ***ap = bp*** iff $a.b = e.f$ and $c.d = g.h$; otherwise ***ap ≠ bp***. Sub-assignment pairs *ap* and *bp* ***intersect***, denoted $ap \cap bp \neq \emptyset$, iff $ap \neq bp$ and *ap* and *bp* can appear together in a candidate assignment, which requires that there are no sibling sub-assignments in $(a.b \cup c.d \cup e.f \cup g.h)$. Clause pair $(c_i, c_j)$ and sub-assignment pair $(i.s, j.t)$ both have ***rank*** $= i + (j(j - 1)/2)$. Clause pairs have total rank order: (0, 1), (0, 2), (1, 2), …, $(m - 2, m - 1)$.

Each sub-assignment pair is ***compatible*** or ***incompatible***. Sub-assignment pair $(i.s, j.t)$ is *compatible* iff (1) the union of literals of *i.s* and *j.t* is non-tautological, and (2) neither *i.s* nor *j.t* is an anticlause; otherwise, $(i.s, j.t)$ is *incompatible*. Every 3-SAT problem in $m$ clauses has $8^2 m(m - 1)/2$ sub-assignment pairs, from the Cartesian product of sub-assignments of $m(m - 1)/2$ clause pairs. Each clause pair corresponds to its $8^2$ sub-assignment pairs.

A ***candidate assignment*** is an ordered set of $m$ sub-assignments, (*i.e.*, an $m$-tuple of sub-assignments), exactly one per clause, with the form: $(\mathbf{0.s_0, 1.s_1, \ldots, (m - 1).s_{m-1}})$, with $0 \leq s_0, s_1, \ldots, s_{m-1} < 8$. Thus, no candidate assignment contains any siblings. A candidate assignment is interpreted as the ordered set of the union of literals in its $m$ sub-assignments. By ${}_mC_2$, the $m$ sub-assignments of each candidate assignment form $m(m - 1)/2$ sub-assignment pairs, from $m(m - 1)/2$ clause pairs. Each 3-SAT problem in $m$ clauses has $8^m$ candidate assignments, the Cartesian product of $m$ 8-ary sub-assignment sets.

## 3 Satisfaction in the Sub-Assignment Expansion

3-SAT satisfaction may be expressed in terms of sub-assignment pairs and candidate assignments.

**Theorem 1.**

*Candidate assignment CA satisfies F iff CA does not contain an incompatible sub-assignment pair.* (3)

*Proof.* Assume $m \geq 2$. If *CA* satisfies *F*, then no sub-assignment of *CA* is an anticlause, (Def., § 2), and no sub-assignment pairs of *CA* are contradictory (inconsistent), (Def., § 2), so that satisfying candidate assignment *CA* contains only compatible sub-assignment pairs. If *CA* does not contain an incompatible sub-assignment pair, then all sub-assignments of *CA* are satisfying non-anticlauses, (Def., § 2), and *CA* is non-contradictory (consistent), so that *CA* is satisfying, by (2). □

**Corollary 1.1.**

*A candidate assignment CA is unsatisfying if and only if CA contains at least one incompatible sub-assignment pair.* (4)

*Proof.* Statement (4) follows by contraposition of each direction of Theorem 1. □

## 4 Each Sub-Assignment Pair Appears in $8^{m-2}$ Candidate Assignments

Because of the mapping of sub-assignment pairs onto the candidate assignments, each sub-assignment pair is contained in (or maps onto) exactly $8^{m-2}$ candidate assignments.

**Theorem 2.**

*Every sub-assignment pair of F appears in exactly* $8^{m-2}$ *candidate assignments of F.* (5)

*Proof.* Let *ip* be a sub-assignment pair. Sub-assignment pair *ip* fixes two sub-assignments for two clauses. The iterated Cartesian product of the remaining ($m$ - 2) 8-*ary* sub-assignment sets generates $8^{m-2}$ ordered combinations of $m$ - 2 sub-assignments, each drawn from $m$ - 2 clauses. Those combine with *ip* to form $8^{m-2}$ candidate assignments. ☐

**Corollary 2.1.**

*Each incompatible sub-assignment pair of F appears in exactly* $8^{m-2}$ *unsatisfying candidate assignments of F.* (6)

*Proof.* Every incompatible sub-assignment pair is a sub-assignment pair. Thus, (6), from Cor. 1.1 and Thm. 2. ☐

## 5 Each Sub-Assignment Pair Appears in a Doubly-Periodic Sub-Structure of $8^{m-2}$ Candidate Assignments of the Search Space

Suppose that the $8^m$ candidate assignments of $F$ are put in some order (specified below), into an $m \times 8^m$ matrix, ***M***. In ***M***, each row $0 \leq i < m$, denoted ***M***[ $i$ ], contains sub-assignments of clause $c_i$, and each column $0 \leq q < 8^m$ one of $8^m$ candidate assignments of $F$. Here I show that each sub-assignment *i.s* is a periodic sub-structure in row ***M***[ $i$ ], and each sub-assignment pair is a doubly-periodic sub-structure of the pair of rows ($i$, $j$) of ***M***. Thus, the locations of all $8^{m-2}$ candidate assignments of each sub-assignment pair are known. Because each incompatible pair is a sub-assignment pair, (def. § 2), and a candidate assignment is unsatisfying if and only if it contains an incompatible sub-assignment pair, (Cor. 1.1), the column locations in ***M*** of all unsatisfying candidate assignments are known *a priori*, once incompatible sub-assignment pairs are determined.

Figures 1-3 show the structures giving rise to the periodically ordered search space described above. (To save space, Figures 1-3 show a 2-SAT problem in three clauses, omitting anticlauses to have only three sub-assignments per clause.) Those structures, used exclusively to prove the periodic interval structure, are:

1. ***DAG*,** a directed acyclic graph of sub-assignments,
2. ***T*,** the 8-ary tree of the preorder traversal of DAG, in which each path is a candidate assignment, and
3. ***M***, the matrix derived from $T$, in which each row, $0 \leq i < m$, denoted ***M***[ $i$ ], contains only the $8^m$ sub-assignments of clause $c_i$, and each column, $0 \leq q < 8^m$, contains a candidate assignment, referenced as ***M***[*][ $q$ ].

Matrix ***M***, used in the proofs below, is generated as follows.

First, let ***DAG*** = {*root*, *N*, *A*} be a directed acyclic graph as normally defined, (see [HRS10]). *Root* is the designated apex node; *N* is the set of $8m$ sub-assignment nodes; and *A* is the set of $8 + 8^2m(m - 1)$ *arcs* connecting certain ordered pairs of nodes. *Root* connects to (child) sub-assignment nodes (0.0, …, 0.7), in that order. For $0 \leq i < (m - 1)$, each $i.s \in$ ($i$.0, ..., $i$.7) connects to all eight (child) sub-assignment nodes in $(i + 1).t \in ((i + 1).0, ..., (i + 1).7)$). (See Fig. 1.)

Second, per [CLR97], a *k-ary* tree is a tree in which root and each non-leaf node have exactly *k* children. (See Fig. 2, showing a 3-ary tree.) Thus, a preorder, depth-first traversal of *DAG* would generate an 8-*ary* tree, ***T***. In *T*, each root to leaf path is an *m*-tuple candidate assignment, i.e., a set containing one sub-assignment per clause. (See Fig. 2.)

Third, while traversing $T$ from leftmost to rightmost path, each path $0 \le q < 8^m$ is placed in $\boldsymbol{M}[*\,][\,q\,]$. (See Fig. 3.) (7)

To formalize that description, let $\otimes$ be the ***repetition operator***, so that $\boldsymbol{r} \otimes \boldsymbol{object}$ denotes $r$ instances of *object*. In the following, each sub-assignment has size 1. For $0 \le i < m$, $0 \le s < 8$, the repeating *objects* and sizes are:

$$\begin{aligned}
\boldsymbol{i.s.block} &= 8^{m-(i+1)} \otimes i.s, \text{ with } 0 \le s < 8, \text{ i.e., } i.s \in (i.0, \ldots, i.7),\\
|\,\boldsymbol{i.s.block}\,| &= \boldsymbol{\beta}_{i.s} = 8^{m-(i+1)};\\
\mathbf{c}_i\boldsymbol{.block} &= (i.0.block, \ldots, i.7.block); \text{ and}\\
|\,\mathbf{c}_i\boldsymbol{.block}\,| &= \boldsymbol{\beta}_{c_i} = 8^{m-i}, \text{ (from } 8 \times |\,i.s.block\,|). \qquad (8)
\end{aligned}$$

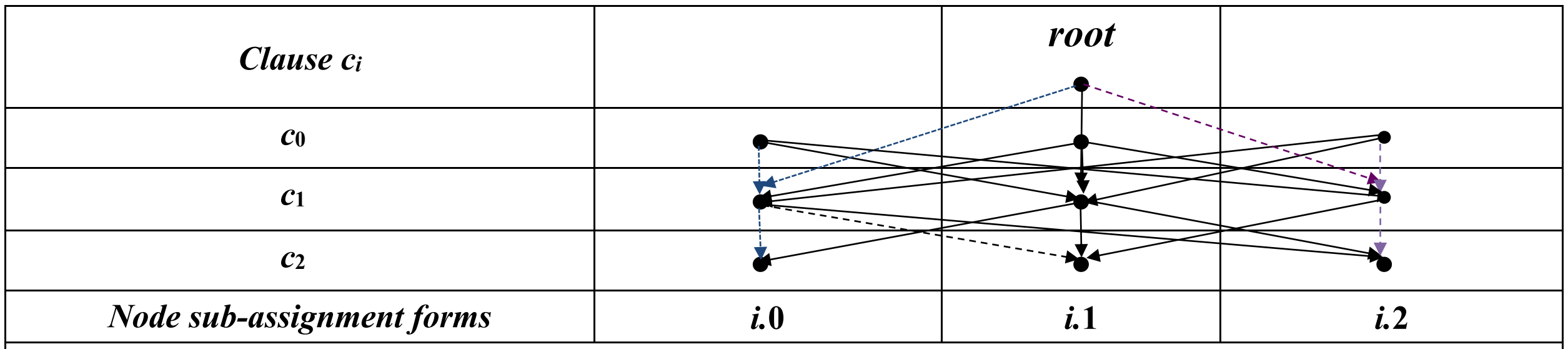

**Figure 1 | Directed acyclic graph of sub-assignment nodes of a 2-SAT problem in three clauses**. With anticlauses omitted to save space, each clause has three sub-assignments. Lines connecting nodes are arcs. Arc direction is from $c_i$ to $c_{i+1}$. Depth- first traversal is from left to right, with all children of node *i.s* visited before siblings are visited, (e.g., visit $path_0$ = <0.0, 1.0, 2.0>; then 2.1, so $path_1$ = <0.0, 1.0, 2.1>; then 2.2, so $path_2$ = <0.0, 1.0, 2.2>, and then 1.1 and 2.0, so $path_3$ = <0.0, 1.1, 2.0>, etc.

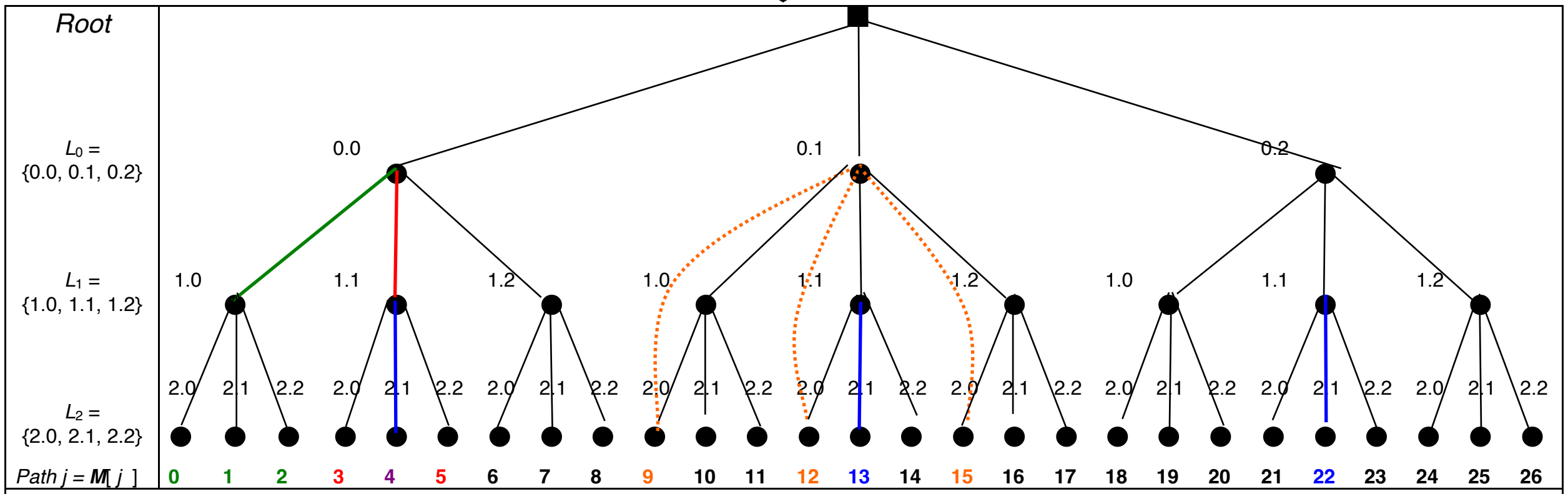

**Figure 2 | The *w*-ary tree, *T*, of lexicographically ordered candidate assignment paths from a preorder depth-first traversal of 2-SAT sub-assignment nodes.** Preorder depth-first traversal tree of 2-SAT problem $F$ with $m = 3$, omitting anticlauses. $F$ has ordered sub-assignment sets $L_0$ = (0.0, 0.1, 0.2), $L_1$ = (1.0, 1.1, 1.2), $L_2$ = (2.0, 2.1, 2.2), each with size $w = 3$. Sub-assignments are nodes in the complete, ordered 3-*ary* tree $T$, with all ordered nodes of level $L_0$ connected by directed edges to all ordered nodes in level $L_1$, which connect to all ordered nodes in level $L_2$. (Fig. 1.) The $j$-th root-to-leaf path from root through nodes of graph levels 0, 1, and 2 is the $j$-th candidate assignment, emitted at the leaf node into matrix column $\boldsymbol{M}[j]$ of Fig. 3. The incompatible sub-assignment pairs (ISAPs) of $F$ are **(0.0, 1.0)**, **(0.0, 1.1)**, **(1.1, 2.1)**, and **(0.1, 2.0)**, shown in the tree by a solid colored line for adjacent sub-assignment pairs, and a dashed line for non-adjacent pairs. Each ISAP renders unsatisfying a periodic set of $w^{m-2} = 3^1 = 3$ of the numbered root-to-leaf paths (shown in the corresponding color). Thus, e.g., **(0.0, 1.1)** causes candidate assignments of paths 3, 4, and 5 to be unsatisfying, and **(1.1, 2.1)** causes candidate assignments of paths 4, 13, and 22 to be unsatisfying. There are satisfying candidate assignments at paths with indices: (6-8, 10, 11, 14, 16-21, 23-26).

**$M$: the $m \times w^m$ Candidate Assignment Matrix has Rows of Periodic Sub-Assignment Blocks, and Doubly-Periodic Sub-Assignment Pairs in Pairs of Rows, Corresponding to Closed Intervals over [0, $w^m$ - 1]**

| # p = $w^d$ | $c_i$ | j = 0 | 1 | 2 | 3 | 4 | 5 | 6 | 7 | 8 | 9 | 10 | 11 | 12 | 13 | 14 | 15 | 16 | 17 | 18 | 19 | 20 | 21 | 22 | 23 | 24 | 25 | 26 |
|---|---|---|---|---|---|---|---|---|---|---|---|---|---|---|---|---|---|---|---|---|---|---|---|---|---|---|---|---|
| 1 | 0 | 0.0 | 0.0 | 0.0 | 0.0 | 0.0 | 0.0 | 0.0 | 0.0 | 0.0 | 0.1 | 0.1 | 0.1 | 0.1 | 0.1 | 0.1 | 0.1 | 0.1 | 0.1 | 0.2 | 0.2 | 0.2 | 0.2 | 0.2 | 0.2 | 0.2 | 0.2 | 0.2 |
| 3 | 1 | 1.0 | 1.0 | 1.0 | 1.1 | 1.1 | 1.1 | 1.2 | 1.2 | 1.2 | 1.0 | 1.0 | 1.0 | 1.1 | 1.1 | 1.1 | 1.2 | 1.2 | 1.2 | 1.0 | 1.0 | 1.0 | 1.1 | 1.1 | 1.1 | 1.2 | 1.2 | 1.2 |
| 9 | 2 | 2.0 | 2.1 | 2.2 | 2.0 | 2.1 | 2.2 | 2.0 | 2.1 | 2.2 | 2.0 | 2.1 | 2.2 | 2.0 | 2.1 | 2.2 | 2.0 | 2.1 | 2.2 | 2.0 | 2.1 | 2.2 | 2.0 | 2.1 | 2.2 | 2.0 | 2.1 | 2.2 |

**Doubly-Periodic Intervals of Incompatible Sub-Assignment Pairs (ISAPs), over Column Indices of M , with Satisfying Regions (SAT)**

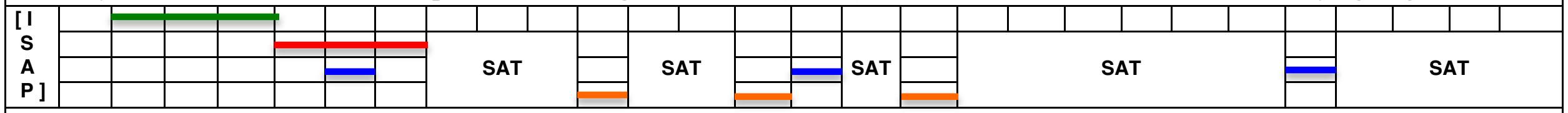

**Figure 3 | $M$, the ordered $m \times w^m$ matrix of candidate assignments for a 2-SAT problem with $m = 3$.** Anticlause sub-assignments are omitted to save space. With $m = 3$ and $w = (2^k - 1) = 3$, there are $w^m = w^3 = 27$ candidate assignments. $M[\ i\ ]$ is the $i$-th row, containing sub-assignments of clause $c_i$. $M[\ *\ ][\ j\ ]$ is the $j$-th column, containing the $j$-th of $w^m$ candidate assignments. For row subscript $0 \le i < m$, row $M[\ i\ ]$ contains $w^i$ instances of a clause block, $c_i.block = (i.0.block, i.1.block, i.2.block)$, where each $i.s.block \in (i.0.block, i.1.block, i.2.block)$ is a contiguous block of $w^{m-(i+1)} = 3^{3-(i+1)}$ $i.s$ sub-assignments. Thus, each $c_i.block$ contains $w^{m-i} = 3^{3-i}$ sub-assignments, over $w^{m-i}$ columns. The leftmost column is the number of instances of $c_i.block$ for clause $c_i$. From Fig. 2, ISAP **(0.0, 1.0)** occurs in one block at indices 0-2, shown in green. ISAP **(1.1, 2.1)** has period 9 and width 1, shown in blue at (4, 13, 22), etc. In [0, 26] no ISAP interval covers a satisfying (SAT) candidate assignment of $M$.

**Theorem 3**. Based on the structures of (8), beginning at column 0 of $M$, for row $0 \le i < m$, the form of row $i$ of $M$ is:

$$\begin{aligned} M[\ i\ ] &= 8^i \otimes (c_i.block), \\ &= 8^i \otimes (i.0.block, ..., i.7.block), \\ &= 8^i \otimes (8^{m-(i+1)} \otimes i.0, \ldots, 8^{m-(i+1)} \otimes i.7). \end{aligned} \qquad (9)$$

*Proof*. Let $T$ be the rooted 8-ary tree generated by the pre-order depth first traversal of $DAG$. (See (7).) In $T$, depth is defined by the clause subscript, $0 \le i < m$. It can be inferred from the $k$-ary tree discussed in [CLR97] that at depth $i$, the set of child nodes $(i.0, ..., i.7)$ is encountered $8^i$ times, and each child node $i.s \in (i.0, ..., i.7)$ is at the head of a set of exactly $8^{m-(i+1)}$ sub-paths, which appear as contiguous paths, in left-to-right order. Thus, $i.s$ appears in a block of $8^{m-(i+1)} \otimes i.s$ contiguous root-to-leaf paths. Generating root-to-leaf paths of $T$ in left-to-right order into $M$ gives row $M[\ i\ ] = 8^i \otimes (8^{m-(i+1)} \otimes i.0, ..., 8^{m-(i+1)} \otimes i.7)$. Substituting terms of (8) proves Theorem 3. □

The column indices of $M$, which are $(0, \ldots, 8^m - 1)$, map one-to-one onto the closed integer interval, $[0, 8^m - 1]$. Thus, each sub-assignment of $F$ may be represented as a periodic interval over $[0, 8^m - 1]$.

**Corollary 3.1 (Periodic intervals of sub-assignments).** The ordered set of column indices of $M$, which is $(0, ..., 8^m - 1)$, maps one-to-one to the closed integer interval $[0, 8^m - 1]$. Let $i.s$ be an arbitrary sub-assignment. Let $\beta_{c_i}, \beta_{i.s}$ be as defined in (8). For $0 \le r_a < 8^i$, sub-assignment $i.s$ appears in the candidate assignments with column locations of $M$ that are described by the periodic interval,

$$\textbf{[}\boldsymbol{i.s}\textbf{]} = [\{\ (r_a \times \beta_{c_i}) + (s \times \beta_{i.s})\ \},\ \{(r_a \times \beta_{c_i}) + (\ (s + 1) \times \beta_{i.s}) - 1)\}].$$

*Proof*. The proof follows by substituting the definitions of (8) into the right-hand side of interval $[i.s]$. □

Theorem 3 suggests several avenues of inquiry, the last of which I pursue below. First, the periodic interval of sub-assignment $i.s$ describes a signal with period $\beta_{c_i}$ (the clause block size), and phase $s$ (the offset into the clause block). Second, the composition of these intervals may be translated to a set of nested "for" loops, potentially subject to dataflow optimization techniques such as [MSKM97] describes. Third, sub-assignment pairs appear as doubly-periodic sub-structures of $M$. That is so because for sub-assignment pair $ap = (i.s, j.t)$, the row $M[\ i\ ]$ columns covered by each repeating block of $i.s$ sub-assignments are the same as the columns covered by $8^{(j-i)-1}$ instances of clause $c_j$ blocks, and each of those $c_j$ blocks contains exactly one $j.t$ block for any fixed $0 \le t < 8$, (by (8)). The following lemmas and theorem prove those propositions.

**Lemma 1.** *The columns of each row $M[\ i\ ]$ i.s.block contain exactly $8^{(j-i)-1} \otimes c_j.block$ instances in row $M[\ j\ ]$.* (10)

*Proof.* Let $0 \leq i < m$ be a row subscript of ***M***. By (8), it can be inferred that in ***M***, the 0-th instance of $c_i$*.block* and $i$.0.*block* begin in column 0. For Statement (10) to be true requires: | *i.s.block* | = $8^{(j-i)-1} \otimes$ | $c_j$*.block* |. Substituting from Statement (8): $8^{m-(i+1)} = 8^{(j-i)-1} \otimes 8^{m-j}$, giving the identity $8^{(j-i)-1} = 8^{(j-i)-1}$, which holds for all repetitions of *i.s.block*, because for fixed $i, j$, block sizes are constant over all repetitions. ☐

**Lemma 2.** *The ratio of i.s.block to $c_j$.block instances is* 1: $8^{(j-i)-1}$.

*Proof.* Lemma 2 follows directly from Lemma 1, because in the columns of each *i.s.block* of row ***M***[ $i$ ], there are exactly $8^{(j-i)-1}$ $c_j$*.block* instances in row ***M***[ $j$ ]. ☐

**Lemma 3.** *For* $0 \leq i < j < m$, *and fixed* $0 \leq s, t < 8$, *the ratio of i.s.block to j.t.block instances is* 1: $8^{(j-i)-1}$.

*Proof.* Because the ratio of *i.s.block to $c_j$.block instances is* 1: $8^{(j-i)-1}$, and for fixed $t$, there is one *j.t.block* in each $c_j$*.block*, Lemma 3 follows. ☐

Consequently, sub-assignment pair (*i.s*, *j.t*) appears as a nested, doubly-periodic sub-structure of ***M***, as Theorem 4 describes.

**Theorem 4 (*Doubly-periodic structure of nested sub-assignment pairs*).** Let *ap* = (*i.s*, *j.t*) be a sub-assignment pair. Let $s$ and $t$ be fixed integers $0 \leq s, t < 8$, so that *i.s.block* ∊ $c_i$*.block* and *j.t.block* ∊ $c_j$*.block* each denote single, unique sub-assignment blocks. Under those definitions,

*the doubly-periodic form of appearances of sub-assignment pair ap in **M** is as **Fig.** 4 depicts*:

***M***[***i***] = { ... |______________________ ***i*.s.*block*** ___________________________|

***i.s.block covers exactly*** $\mathbf{8^{(j-i)-1}} \otimes$ ***$c_j$.block*, *where***

...

$\mathbf{8^{(j-i)-1}} \otimes$ ***$c_j$.block***

=

***M***[***j***]= { ... |_______ ***$c_j$.block*[0]** _______|, ..., |____ ***$c_j$.block*[($\mathbf{8^{(j-i)-1}}$) - 1]** ____|

...

=

... |_ (***j*.0.*block*, ..., *j*.7.*block***) __|, ..., |___ (***j*.0.*block* , ..., *j*.7.*block***) ___|

**Figure 4 | Nested form of doubly-periodic repetitions of sub-assignment pair.**

*Proof.* By Lemma 2, each instance of *i.s.block* in row ***M***[*i*] covers the same columns as $8^{(j-i)-1} \otimes$ $c_j$*.block* instances *in row **M**[j]*. Each $c_j$*.block* contains eight sub-assignment blocks, (by Stmt. 8), exactly one of which is *j.t.block*, for any fixed $0 \leq t < 8$. ☐

Combining Theorems 3 and 4 gives the exact column locations in ***M*** of each sub-assignment pair of $F$, which can be represented as a doubly-periodic interval. Thus, once all the incompatible sub-assignment pairs of $F$ have been identified, the search space locations of all unsatisfying candidate assignments of $F$ are known.

**Theorem 5 (*Column indices of doubly-periodic sub-assignment pairs*).** Let *ap* = (*i.s*, *j.t*) be a sub-assignment pair, with fixed $i, j$, $s$, and $t$. Let $\beta_{c_i}$= | $c_i$*.block* | = $8^{m-i}$, and let $\beta_{i.s}$ = | *i.s.block* |. (See (8).) Sub-assignment pair *ap* appears in the following family of $8^i$ families of $8^{(j-i)-1}$ sets of locations of ***M***:

$$\bigcup_{0 \leq r_i < 8^i} \; \bigcup_{0 \leq r_{ij} < 8^{(j-i)-1}} \{left_{ij}, \ldots, right_{ij}\},$$

*where*: $b_i \quad = (r_i \times \beta_{c_i}) + (s \times \beta_{i.s})$

$$= 8^{m-i}(r_i + \tfrac{s}{8}),$$

$$left_{ij} = b_i + (r_{ij} \times \beta_{c_j}) + (t \times \beta_{j.t})$$

$$= b_i + 8^{m-j}(r_{ij} + \tfrac{t}{8}), \text{ and}$$

$$right_{ij} = (left_{ij} + \beta_{j.t}) - 1. \quad (11)$$

*Proof.* The proof follows by substituting the block sizes of Statement (8) into the ratios of Lemma 2. ☐

Because sub-assignment pair $ap = (i.s, j.t)$ occurs only in columns where both $i.s$ and $j.t$ coincide,

$$|\ ap.block\ | = |\ j.t.block\ |. \quad (12)$$

Because each sub-assignment is periodic and each sub-assignment pair is doubly-periodic in the candidate assignment search space, an algorithm need not represent graph *DAG*, tree *T*, or matrix ***M***. Instead, as the next section shows, an algorithm may represent all $8^m$ candidate assignments of *F* using no more than $8^2m(m - 1)/2$ doubly-periodic closed integer intervals over the closed integer $[0, 8^m - 1]$.

## 6 The Set of $8^{m-2}$ Candidate Assignments of Each Sub-Assignment Pair is a Doubly-Periodic Interval over the Closed Integer Interval, $[0, 8^m - 1]$

If the column indices of ***M***, which are $(0, ..., 8^m - 1)$, are mapped one-to-one to the closed integer interval $\mathbf{[0, 8^m - 1]}$, then a doubly-periodic interval over closed integer interval $\mathbf{[0, 8^m - 1]}$ may be used to describe the set of column locations in ***M*** of each sub-assignment pair, as Corollary 5.1 describes. Therefore, all unsatisfying candidate assignments of a 3-SAT problem *F* in *m* clauses may be described using a set of no more than $8^2m(m - 1)/2$ doubly-periodic intervals. Those intervals give the locations of all unsatisfying candidate assignments of *F*.

**Corollary 5.1.** The ***doubly-periodic interval of sub-assignment pair ap*** $= (i.s, j.t)$ is:

$$[[ap]] = (\ [\ l(r_i, i.s) + l(r_{ij}, j.t),\ l(r_i, i.s) + l(r_{ij}, j.t) + |\ j.t.block\ | - 1]\ ), \text{ where}$$
$$l(r_i, i.s) = (r_i \times |\ c_i.block\ |) + (s \times |\ i.s.block\ |),$$
$$l(r_{ij}, j.t) = (r_{ij} \times |\ c_j.block\ | + (t \times |\ j.t.block\ |), \text{ and}$$
$$0 \le r_i < 8^i, 0 \le r_{ij} < 8^{(j-i)-1},$$

which reduces to:

$$[[ap]] = (\ [8^{m-i}(r_i + s/8) + 8^{m-j}(r_{ij} + t/8),\ (8^{m-i}(r_i + s/8) + 8^{m-j}(r_{ij} + (t+1)/8)) - 1]\ ),$$
$$\text{where } 0 \le r_i < 8^i \text{ and } 0 \le r_{ij} < 8^{(j-i)-1}. \quad (13)$$

*Proof.* The proof is by substitution from (11). For 2-SAT, with $w = 2^2$, instead of $w = 2^3$ as for 3-SAT, substitute 4 for 8 in (13). ☐

**Corollary 5.2.**

*The search space locations of all unsatisfying candidate assignments of F are known a priori, once the incompatible sub-assignment pairs of F are determined.*

*Proof.* Each sub-assignment pair *ap* determines the exact set of search space locations of all $8^{m-2}$ candidate assignments containing *ap*. (Thm. 5.) A candidate assignment is unsatisfying iff it contains an incompatible sub-assignment pair. (Cor. 1.1.) Thus, each incompatible sub-assignment pair *ip* determines the exact set of search space locations of all $8^{m-2}$ unsatisfying candidate assignments containing *ip*, which are known once the incompatible sub-assignment pairs of *F* are known. The union of all such search space locations is exactly the set of locations of all unsatisfying candidate assignments of *F*, (Cor. 1.1), so that Corollary 5.2 follows. ☐

## 7 In a System of Lights, Only Unsatisfying Candidate Assignments are Shadowed

Every 3-SAT problem models a system of lights and perfect shadows. Suppose that $8^m$ lights, indexed left-to-right over the closed integer interval $[0, 8^m - 1]$ run above a set of no more than $8^2m(m - 1)/2$ vertically arranged tracks. Each track has a doubly-periodic interval of an incompatible sub-assignment pair. (See Fig. 5.) Each interval of an incompatible sub-assignment pair *ip* covers the indices of the candidate assignments in which *ip* appears. (See (11) and (13).) The floor beneath the light and interval tracks contains a set of $8^m$ candidate assignments, indexed [0, $8^m$ - 1]. Every point *p* in incompatible sub-assignment pair interval [[*ip*]] casts a shadow onto either (1) the corresponding point *p* of any intersecting incompatible sub-assignment pair intervals beneath [[*ip*]], or (2) onto the candidate assignment beneath *p*, but (3) not onto any other point. In that system, all shadowed candidate assignments are unsatisfying; but all un-shadowed candidate assignments are satisfying. The union of periodically shadowed points is the set of unsatisfying candidate assignments of *F*. Let *ip* and *jp* be incompatible sub-assignment pairs. Shadowing is defined formally as follows:

1. *ip* ***shadows*** (and is a shadower of) the $8^{m-2}$ points directly under [[*ip*]], (i.e., the unsatisfying candidate assignment indices of *ip*), but *ip* shadows no other points.
2. *ip* ***shadows*** *jp* if and only if (1) *ip.rank* < *jp.rank*, and (2) *ip* intersects *jp*.
3. No compatible sub-assignment pair shadows anything.
4. There are no other shadows.
5. *ip* is an ***unshadowed shadower*** only over the points over which *ip* is not shadowed. (14)

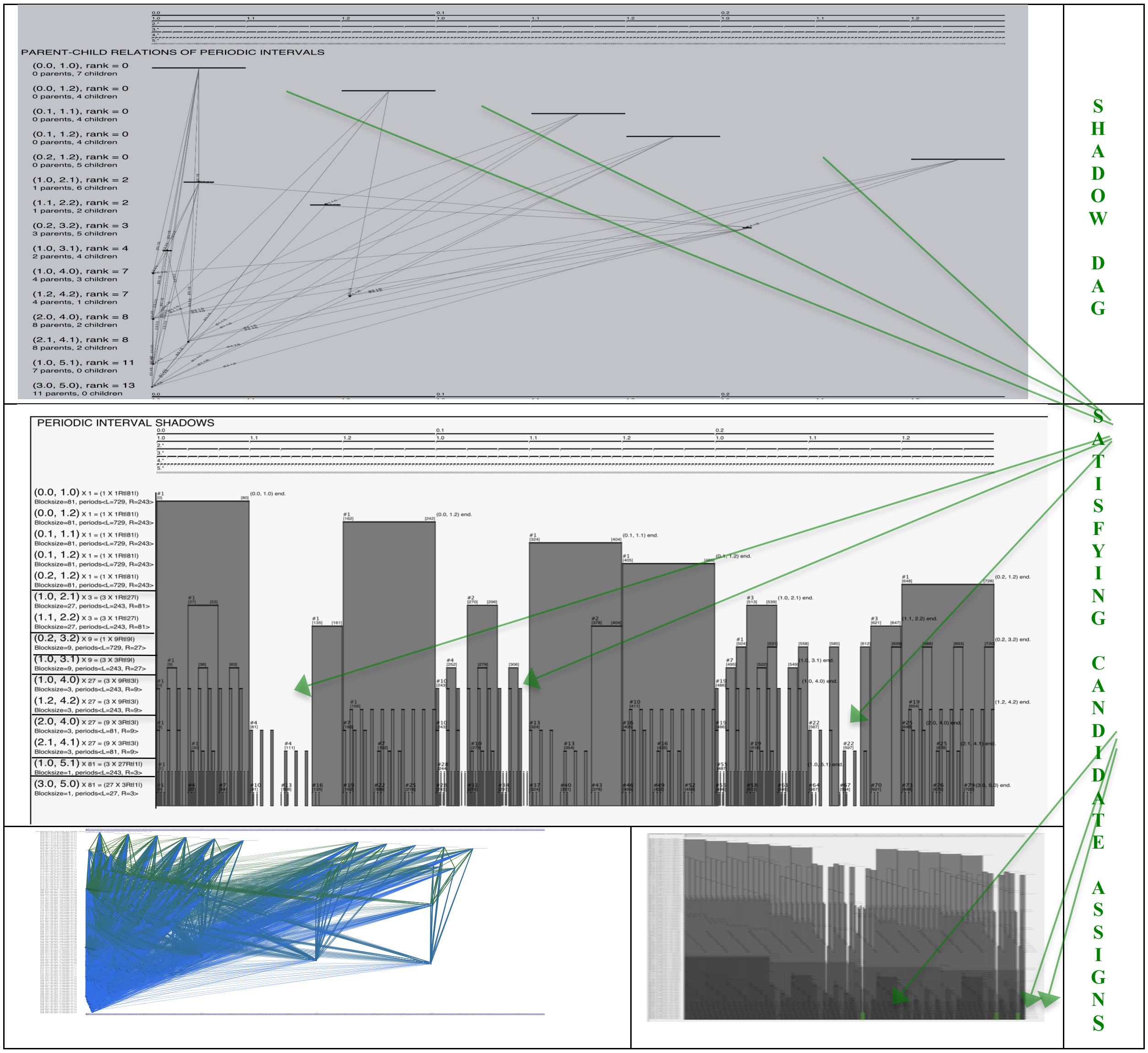

**Figure 5. Doubly-Periodic Intervals of Incompatible Sub-Assignment Pairs.** Left columns label incompatible sub-assignment pair intervals (ISAPs), grouped by rank, with ranks divided by a solid vertical line. The top scale is the run of sub-assignments over $[0, w^m - 1]$. **Top figure**: 2-SAT problem *F* with $m = 6$ and $w = 3$, (anticlauses omitted), showing rank-ordered directed acyclic graph of shadowing ("ISAP"). The first instance of each ISAP is to the right of each label; an arc connects each ISAP to each ISAP it shadows. **Middle figure**: Complex, nested shadow patterns arise when each ISAP of the top figure is extended over its closed integer interval range $[0, 3^6 - 1]$. The grey rectangle beneath each interval shows the shadowed region. Regions: Each sub-assignment pair of a rank covers $1/w^2 = 1/9$ of the candidate assignments; the $w^2$ sub-assignments of a rank cover all candidate assignments. Rank, Frequency, & Block Width: For ISAP $ip = (i.s, j.t)$, *i* determines *i.s.block* width and frequency, and *s* the relative position (phase) of *i.s.block* in each $c_i$*.block*; $w^{(j-i)-1}$ gives frequency of $c_j$*.block* within each *i.s.block*, and *t* gives relative position (phase) of *j.t.block* in each $c_j$*.block*. Increasing rank or right clause subscript increases frequency but decreases block width. (Compare (0.0, 1.0), (0.2, 3.2), and (1.0, 3.1).) Shadowed are Unsatisfying: There are no satisfying candidate assignments in ISAP-shadowed regions. Satisfying in Lowest Density & Least Covered: Comparing the region under compatible sub-assignment pair (CSAP), (0.0, 1.1), to those under CSAPs (0.1, 1.0), (0.2, 1.0) and (0.2, 1.1), it appears that ISAP density corresponds inversely to satisfying candidate assignment locations in a region. Counting Unique Unsatisfying Candidate Assignments: Summing points under which each ISAP interval is an ***unshadowed shadower*** gives the number of unsatisfying candidate assignments of *F*. E.g., under CSAP (0.0, 1.0), ISAP (1.1, 2.2) is an unshadowed shadower covering 27 unsatisfying candidate assignments; but under ISAP (0.1, 1.1), (1.1, 2.2) is not an unshadowed shadower, but is shadowed. Periodic sub-structures indicate state-augmented dynamic programming may apply to 3-SAT. **Bottom figure:** Base interval DAG (left) and search space (right) for the simple 2-SAT problem, ($m = 5$, $w = 4$, and 1024 candidate assignments), of **[APT79]**. As in hard 3-SAT problems, few (only three) of the candidate assignments are satisfying. Increasing *m* magnifies those (not shown).

The incompatible sub-assignment pairs of 3-SAT problem *F* may be ordered into a graph of (not-necessarily-transitive) shadow relations. The shadow relations of incompatible sub-assignment pairs form a directed acyclic (shadow) graph, ***SG* = (*N*, *A*)**. Each node $ip \in N$ is an incompatible sub-assignment pair of *F*. *A* is the set of arcs describing shadow relations of pairs of nodes of *N*, as follows. Node *ip* is associated with two possibly empty families of sets, ***ip.parents*** and ***ip.children***. The family ***ip.parents*** contains the set of incompatible sub-assignment pairs that shadow *ip*, in increasing rank order, so that for $0 \le r < ip.rank$, ***ip.parents*[ *r* ]** is the (possibly empty) set of incompatible sub-assignment pairs of rank *r* that shadow *ip*. For $0 \le ip.rank < r$, ***ip.children*[ *r* ]** is the set of incompatible sub-assignment pairs of rank *r* that *ip* shadows. For nodes $ip, jp \in N$, a directed arc $a \in A$ connects a node *ip* to *jp* iff $jp \in ip.children$; i.e., iff *ip* shadows *jp*. Incompatible sub-assignment pair $jp \in ip.children$ is ***nested*** node (or a nested shadow) iff a parent of *ip* shadows *jp*, i.e., iff:

$$\exists\, hp \in ip.parents\ (\, jp \in hp.children) \wedge (jp \in ip.children)\ ), \text{ i.e.,}$$
$$\exists\, hp \in ip.parents\ (\, jp \in (hp.children \wedge ip.children)\ ). \tag{15}$$

For example, in the top sub-figure of Figure 5, (2.0, 4.0) is nested under (0.2, 1.2), (1.0, 3.1), and (1.0, 4.0).

**Theorem 6.** *In the system of lights and shadowing intervals described in* (14), *above, no satisfying candidate assignment is shadowed, but every unsatisfying candidate assignment is shadowed.*

*Proof.* Because each incompatible sub-assignment pair shadows only unsatisfying candidate assignments, by (14), no satisfying candidate assignment is shadowed. Every unsatisfying candidate assignment *u* contains at least one incompatible sub-assignment pair, (Thm. 1), which shadows *u*, by (14). Thus, every unsatisfying candidate assignment is shadowed. ☐

**Corollary 6.1.** Let *F* be a 3-SAT problem in *m* clauses, with *IP* the set of all | *IP* | incompatible sub-assignment pairs of *F*, and *U* the set of all unsatisfying candidate assignments of *F*. For $ip \in IP$, let ***ip.hi*** be the set of unsatisfying candidate assignments over which *ip* is an unshadowed shadower, and let ***ip.un*** = | *ip.hi* |. Then,

$$|U| = \sum_{ip\in IP} ip.un = \sum_{ip\in IP} |\, ip.hi\,| = \left|\bigcup_{ip\in IP} ip.hi\right|. \tag{16}$$

*Proof.* Proof is of the leftmost equality. Each $u \in U$ contains at least one incompatible sub-assignment pair, (Cor. 1.1). Only one of the $m(m - 1)/2$ rank-ordered sub-assignment pairs of *u* is the lowest-ranked incompatible sub-assignment pair of *u*, (§ 2, *rank* def.), i.e., an unshadowed shadower. If *ip* is lowest ranked in *u*, *ip.un* counts only *ip* for *u*, i.e., each candidate assignment under an unshadowed shadower is counted just once. Thus, Corollary 6.1. ☐

## 8 Inclusion-Exclusion over $m^3$ Coefficients Calculates the Number of Unsatisfying Candidate Assignments

The best known upper-bound time complexity to count the number of satisfying candidate assignments of a 3-SAT problem *F* in *n* variables and *m* clauses is $O^*(1.637^n)$ ([KUT26]; see also [KUT07] $O(1.6423^n)$ and [DVMPJW2005] $O(1.6737^n)$, on model counting). Other authors have explored exact and approximate inclusion-exclusion, including as applied to 3-SAT model counting. (See [IWA89], [MASS97], [KJLNSA96], [LNNN90].) Here, I show that the number of unsatisfying candidate assignments of a 3-SAT problem can be calculated in $O(m^3 + \mathit{Oracle})$ time, using the inclusion-exclusion formula. *Oracle* is the cost to calculate the number of intersections of (2, …, $m(m - 1)/2$) incompatible sub-assignment pairs containing (3, …, *m*) sub-assignments. (Unlike the Delphic oracle, this one is unlikely to be cheap to consult.) The calculation rests on the inclusion-exclusion principle, which provides: the cardinality of the union of a family of sets, $A = \{A_1 ... A_q\}$ is an alternating sum of cardinalities of intersections of increasingly larger subfamilies of *A*. Under inclusion-exclusion, ([HKS03]), the cardinality of the union of the members of *A* is the alternating sum of $|A|$ series, so that:

$$
\begin{aligned}
|A_1 \cup A_2 \cup ... \cup A_q| = \; & \textstyle\sum_{1 \le i \le q} |A_i| && (e_1)\\
& - \textstyle\sum_{1 \le i < j \le q} |A_i \cap A_j| && (e_2)\\
& + \textstyle\sum_{1 \le i < j < k \le q} |A_i \cap A_j \cap A_k| && (e_3)\\
& - ... \\
& + (-1)^{q+1} \textstyle\sum_{1 \le i < j < k ... \le q} |A_i \cap A_j \cap A_k \cap ... \cap A_q|, && (e_q)\\
& \text{where } q = |A|. && (17)
\end{aligned}
$$

### 8.1 3-SAT INCLUSION-EXCLUSION

For 3-SAT, each $A_i$ term of (17) corresponds to the set of unsatisfying candidate assignments of an incompatible sub-assignment pair. Let *IP* be the set of incompatible sub-assignment pairs of 3-SAT problem *F*, and let $ip_i \in IP$ be an incompatible sub-assignment pair that appears in the $8^{m-2}$ unsatisfying candidate assignments containing $ip_i$, denoted $ip_i.u$. In line $e_1$ of (17), each $|A_i| = |ip_i.u|$. Likewise, in line $e_2$, each $|A_i \cap A_j|$ corresponds to a pair of intersecting incompatible sub-assignment pairs, $(ip_i \cap ip_j)$, which appear together in $|ip_i.u \cap ip_j.u|$ unsatisfying candidate assignments, where $ip_i, ip_j \in IP,$ and so on for lines $e_3$ through $e_q$, so that 3-SAT inclusion-exclusion has the form:

$$
\begin{aligned}
|ip_1.u \cup ... \cup ip_q.u| = \; & \textstyle\sum_{1 \le i \le q} |ip_i.u| && (e_1)\\
& - \textstyle\sum_{1 \le i < j \le q} |ip_i.u \cap ip_j.u| && (e_2)\\
& + \textstyle\sum_{1 \le i < j < k \le q} |ip_i.u \cap ip_j.u \cap ip_k.u| && (e_3)\\
& - ... \\
& + (-1)^{q+1} \textstyle\sum_{1 \le i < j < k < ... \le q} |ip_i.u \cap ip_j.u \cap ... \cap ip_q.u|, && (e_q)\\
& \text{where } q = |A|. && (18)
\end{aligned}
$$

Each line $e_i$ in (18) corresponds to a range of possible cardinalities. Let $ip_i.left$ and $ip_i.right$ be the left and right sub-assignments of incompatible sub-assignment pair $ip_i$. For line $e_1$, $|ip_i.u| = 8^{m-2}$, (Thm. 2), so that line $e_1$ is $|IP| \times 8^{m-2}$. For line $e_2$, if $ip_i$ and $ip_j$ intersect, then $S = ip_i.left \cup ip_i.right \cup ip_j.left \cup ip_j.right$. By the definition of intersection (see § 2), $|S| \in \{3, 4\}$. Thus for each intersection in line $e_2$, $|ip_i \cap ip_j| \in \{8^{m-3}, 8^{m-4}\}$. Likewise, each intersection of *e* incompatible sub-assignment pairs contains *s* sub-assignments, where $s \in (smin_e, …, smax_e)$ is an ***s*-category** for *e*. As the following lemmas and theorem show, if an oracle can determine the number of such *s*-categories for *e*, denoted $\boldsymbol{n_{s(e)}}$, then 3-SAT inclusion-exclusion can be calculated in no more than $m^3$ multiplications (excluding the non-trivial cost of *Oracle*). Lemmas 4 and 5 give the range of the number of sub-assignments and corresponding unsatisfying candidate assignments of a set of *e* mutually intersecting incompatible sub-assignment pairs. Theorem 7 gives the derived form of coefficient-based inclusion-exclusion for 3-SAT.

**Lemma 4.** Let $F$ be a 3-SAT problem in $m$ clauses. Let $ip$ be a sub-assignment pair with left and right member sub-assignments $ip.l$ and $ip.r$. Let $I_e$ be a set of $2 \leq e \leq m(m - 1)/2$ mutually intersecting sub-assignment pairs of $F$, and let $S = \bigcup_{ip \in I_e}(ip.l \cup ip.r)$ be the set of sub-assignments of $I_e$. Under those definitions, $|S| \in \{smin_e \dots smax_e\}$ where:

$$smin_e = ceiling(\frac{1+\sqrt{1+8e}}{2}), \text{ and}$$
$$smax_e = \min(2e, m). \tag{19}$$

*Proof.* By ${}_mC_2$, $s$ sub-assignments generate at most $s(s - 1)/2$ sub-assignment pairs, so that the $e$ sub-assignment pairs of $I_e$ require $s(s - 1)/2 \geq e$. Thus, $s^2 - s - 2e \geq 0$, from which, $smin_e = \frac{1+\sqrt{1+8e}}{2}$, (by q.f.), restricted by *ceiling* because $smin_e$ is an integer. For $smax_e$, (1) the $e$ sub-assignment pairs of $I_e$ cover no more than $2e$ sub-assignments, because each sub-assignment pair has two sub-assignments, and (2) if $2e > m$, then $smax_e = m$, since all members of $I_e$ are mutually intersecting and by the definition of $I_e$, $|S| \leq m$, (§ 2, cand. assign. def.). Thus, Lemma 4. ☐

**Lemma 5.** Reusing the definitions of Lemma 4, let $I_e$ be a set of $2 \leq e \leq m(m - 1)/2$ mutually intersecting sub-assignment pairs of $F$, and let $smin_e$ and $smax_e$ be defined as in (19) of Lemma 4. The members of $I_e$ appear together in $z$ candidate assignments, where $z \in \{8^{m-smin_e}, \dots, 8^{m-smax_e}\}$.

*Proof.* From Theorem 2, a set of $|S|$ non-sibling sub-assignments appears in $8^{m-|S|}$ candidate assignments. Thus, with $|S| \in \{smin_e \dots smax_e\}$, (by Lemma 4), $z \in \{8^{m-smin_e}, \dots, 8^{m-smax_e}\}$. ☐

Let $F$ be 3-SAT problem in $m$ clauses, with set of incompatible sub-assignment pairs, $IP$. For $2 < e \leq m(m - 1)/2$, let $I_e$ be an arbitrary set of $e$ mutually intersecting incompatible sub-assignment pairs $ip_1, \dots, ip_e \in IP$. Let $IC_e$ be the set of all such mutually intersecting combinations of $e$ incompatible sub-assignment pairs of $IP$. For arbitrary sub-assignment pair $ip \in IP$, denote the left and right sub-assignments $ip.l$ and $ip.r$. By Lemma 4, the union of sub-assignments of any $I_e \in IC_e$ contains between $smin_e \dots smax_e$ sub-assignments. Let coefficient $\boldsymbol{n}_{s(e)}$ be the number of members of $I_e \in IC_e$ such that the union of members of $I_e$ contains $s$ sub-assignments, i.e.:

$$\boldsymbol{n}_{s(e)} = |\,(\, I_e \in IC_e : \, |\bigcup_{ip \in I_e} ip.l \cup ip.r| = s\,)\,|. \tag{20}$$

**Theorem 7.** Suppose that an oracle has calculated all of the $\boldsymbol{n}_{s(e)}$ coefficients for all sets of between 2 and $m(m - 1)/2$ (inclusive) intersecting incompatible sub-assignment pairs of $F$, i.e., for all $e_2 \dots e_{m(m-1)/2}$. Let $y = m(m - 1)/2$ and let $U$ be the set of unique unsatisfying candidate assignments of 3-SAT problem $F$, so that $|U|$ is the number of unsatisfying candidate assignments of $F$. The following equation counts the number of unique unsatisfying candidate assignments of $F$.

$$
\begin{aligned}
|U| = \; & |IP| \times 8^{m-2} && (e_1)\\
& - (\, (n_{3(2)} \times 8^{m-3}) + (n_{4(2)} \times 8^{m-4})\,) && (e_2)\\
& + ((n_{3(3)} \times 8^{m-3}) + (n_{4(3)} \times 8^{m-4}) + (n_{5(3)} \times 8^{m-5}) + (n_{6(3)} \times 8^{m-6})) && (e_3)\\
& - \dots && (e_{\dots})\\
& + (-1)^{y+1} \times ((nsmin_{(y)} \times 8^{m-smin_y}) + \dots + (nsmax_{(y)} \times 8^{m-smax_y}), && (e_y)\\
& \text{with } y = m(m - 1)/2. && (21)
\end{aligned}
$$

*Proof.* Let $IP$ be the set of incompatible sub-assignment pairs of a 3-SAT problem $F$. For $ip_i \in IP$, let $ip_i.u$ be the set of $8^{m-2}$ unsatisfying candidate assignments containing incompatible sub-assignment pair $ip_i$; let $|ip_i| = |ip_i.u|$ be the number of unsatisfying candidate assignments containing $ip_i$, i.e., $8^{m-2}$. Let $A = (ip_1.u, \dots, ip_q.u)$ be a family of $q$ sets of unsatisfying candidate assignments of the incompatible sub-assignment pairs of $F$. The inclusion-exclusion rule of the equation in (17) gives the number of unsatisfying candidate assignments of $F$, which is:

$$
\begin{aligned}
|ip_1.u \cup \dots \cup ip_q.u| = \; & \sum_{1 \leq i \leq q} |A_i| && (e_1)\\
& - \sum_{1 \leq i < j \leq q} |A_i \cap A_j| && (e_2)\\
& + \sum_{1 \leq i < j < k \leq q} |A_i \cap A_j \cap A_k| && (e_3)\\
& - \dots \\
& + (-1)^{q+1} \sum_{1 \leq i < j < k \dots \leq q} |A_i \cap A_j \cap A_k \cap \dots \cap A_q|, && (e_q)
\end{aligned}
$$

$$\text{where } q = |A|. \tag{22}$$

Non-intersecting sets of incompatible sub-assignment pairs that are visited in (22) cover zero candidate assignments (§ 2, cand. assign. def.); those do not affect the sum. Because $A = (ip_1.u, ..., ip_q.u)$, the right side of lines $e_1$ through $e_q$ of equation (22) are:

$$
\begin{aligned}
|ip_1.u \cup ... \cup ip_q.u| = \quad & \textstyle\sum_{1 \le i \le q} |ip_i.u| && (e_1)\\
& - \textstyle\sum_{1 \le i < j \le q} |ip_i.u \cap ip_j.u| && (e_2)\\
& + \textstyle\sum_{1 \le i < j < k \le q} |ip_i.u \cap ip_j.u \cap ip_k.u| && (e_3)\\
& - ...\\
& + (-1)^{q+1} \textstyle\sum_{1 \le i < j < k < ... \le q} |ip_i.u \cap ip_j.u \cap ... \cap ip_q.u|. && (e_q)\\
& \text{where } q = |A| = |IP|. && (23)
\end{aligned}
$$

By Lemma 4, any set of $1 \le z \le m(m - 1)/2$ mutually intersecting incompatible sub-assignment pairs such as $(ip_i.u \cap ip_j.u \cap ... \cap ip_z.u)$ covers $s = |(ip_i \cup ip_j \cup ... \cup ip_z)|$ sub-assignments such that $s \in \{smin_z, ..., smax_z\}$. By Lemma 5, any intersection of $z$ incompatible sub-assignment pairs such as $(ip_i \cap ip_j \cap ... \cap ip_z)$ covers $p \in (8^{m - smin_z}, ..., 8^{m - smax_z})$ unsatisfying candidate assignments. Thus, the cardinality calculations in lines $e_1$ through $e_q$ of equation (23) correspond to the identities:

$$
\begin{aligned}
&(e_1):\ |ip_i.u| && = 8^{m-2};\\
&(e_2):\ |ip_i.u \cap ip_j.u| && = 8^{m-|S|} : 3 \le |S| \le 4; \text{ and}\\
&(e_3):\ |ip_i.u \cap ip_j.u \cap ip_k.u| && = 8^{m-|S|} : 3 \le |S| \le 6,\\
& && ...\\
&(e_y):\ |ip_i.u \cap ip_j.u \cap ip_k.u \cap ... \cap ip_q.u| && = 8^{m-|S|} : smin_y \le |S| \le smax_y,\\
& && \text{where } |(ip_i, ip_j, ip_k, ..., ip_q)| = y. \qquad (24)
\end{aligned}
$$

According to the definition in (20), for each $s$-category of $8^{m - |S|}$, there are $\boldsymbol{n_{s(e)}}$ intersections of $e$ incompatible sub-assignment pairs containing $s$ sub-assignments. Thus, the right side summations of the equation in (23) are the identities:

$$
\begin{aligned}
&\textstyle\sum_{1 \le i \le q} |ip_i.u| && = |IP| \times 8^{m-2} && (e_1)\\
&\textstyle\sum_{1 \le i < j \le q} |ip_i.u \cap ip_j.u| && = (n_{3(2)} \times 8^{m-3}) + (n_{4(2)} \times 8^{m-4}) && (e_2)\\
&\textstyle\sum_{1 \le i < j < k \le q} |ip_i.u \cap ip_j.u \cap ip_k.u| && = (n_{3(3)} \times 8^{m-3}) + (n_{4(3)} \times 8^{m-4}) + (n_{5(3)} \times 8^{m-5}) + (n_{6(3)} \times 8^{m-6}) && (e_3)\\
& && ...\\
&\textstyle\sum_{1 \le i < ... \le q} |ip_i.u \cap ... \cap ip_q.u| && = (nsmin_{(y)} \times 8^{m - smin_y}) + ... + (nsmax_{(y)} \times 8^{m - smax_y}), && (e_y)\\
& && \text{where } q = |A| = |IP| \text{ and } |(ip_i, ip_j, ip_k, ..., ip_q)| = y. && (25)
\end{aligned}
$$

Note that (22) and (23) account for zero-valued intersections, including intersections of more than $m(m-1)/2$ incompatible sub-assignment pairs and intersections of ISAPs containing sibling sub-assignments, which cannot appear in candidate assignments. Next, we again consider $s$-categories of sub-assignment cardinalities of intersections of between $(1, ..., m(m - 1)/2)$ incompatible sub-assignment pairs contained in unsatisfying candidate assignments. Thus, without loss of generality, we revert to bounding the length of the last nonzero term by $y = m(m - 1)/2$, as in Thm. 7 (stmt. (21)). Substituting the right side of the equations of (25) into the right side of (23) gives Theorem 7:

$$
\begin{aligned}
|U| = \ & |IP| \times 8^{m-2} && (e_1)\\
& - ((n_{3(2)} \times 8^{m-3}) + (n_{4(2)} \times 8^{m-4})) && (e_2)\\
& + ((n_{3(3)} \times 8^{m-3}) + (n_{4(3)} \times 8^{m-4}) + (n_{5(3)} \times 8^{m-5}) + (n_{6(3)} \times 8^{m-6})) && (e_3)\\
& - ...\\
& + (-1)^{y+1} \times ((nsmin_{(y)} \times 8^{m - smin_y}) + ... + (nsmax_{(y)} \times 8^{m - smax_y}), && (e_y)\\
& \text{with } y = m(m - 1)/2. && \square
\end{aligned}
$$

**Corollary 7.1.** The number of unsatisfying candidate assignments of any 3-SAT problem $F$ in $m$ clauses is:

$$| U | = | IP | \times 8^{m-2} \quad (e_1)$$
$$+ (-1)^{g+1} \times ((nsmin_{(g)} \times 8^{m - smin_g}) + ... + (nsmax_{(g)} \times 8^{m - smax_g}), \quad (e_g)$$
where the $(e_g)$ term is summed over $2 \le g \le m(m - 1)/2$. (26)

*Proof.* By substitution from Lemmas 4 and 5 for the $e_2$ through $e_y$ terms of Theorem 7. □

## 8.2 3-SAT INCLUSION-EXCLUSION HAS $O$( ORACLE + $M^3$ ) COMPLEXITY

Assuming an oracle has determined the $O(m^3)$ $\boldsymbol{n_{s(e)}}$ coefficients required by Theorem 7—and excluding the cost of that oracular determination—Theorem 7 calculates the number of unsatisfying candidate assignments of 3-SAT problem $F$ using fewer than $m^3$ multiplications.

**Lemma 6.** *The coefficient form of inclusion-exclusion of Theorem 7 calculates the number of unsatisfying candidate assignments in fewer than $m^3$ multiplications.*

*Proof.* Let $I_e$ be an arbitrary set of $e$ intersecting incompatible sub-assignment pairs, and let $S$ be the union of sub-assignments of $I_e$. By Lemma 4, $| S | \in \{smin_e, ..., smax_e\}$. By Lemma 5, the $| S |$ sub-assignments of $S$ appear in a set of $8^{m - |S|}$ unsatisfying candidate assignments. Where $min = smin_e$, and $max = smax_e$, an intersection of $e$ sub-assignment pairs covers between $8^{m - min}$ ... $8^{m - max}$ unsatisfying candidate assignments. Thus, each line of Theorem 7 performs no more than $m$ multiplications, because, for $2 \le e \le m(m - 1)/2$, it follows that $smax_e - smin_e < m$. Consequently, the no more than $m(m - 1)/2$ lines of Theorem 7 perform no more than $m^3$ total multiplications. □

**Theorem 8.** *If an oracle calculates the $\boldsymbol{n_{s(e)}}$ coefficients of 3-SAT problem F in time Oracle, then calculating the number of unsatisfying candidate assignments of F takes* $O(Oracle + m^3)$ time.

*Proof.* Once the $\boldsymbol{n_{s(e)}}$ *coefficients* are determined in time, *Oracle*, the cardinality calculation in Theorem 7 requires $O(m^3)$ multiplications. Thus, Theorem 8 holds. □

# 9 Recurse on Shadows to Calculate Unique Unsatisfying Candidate Assignments

The equation in Theorem 7 calculates, breadth-first, the nested shadows of each incompatible sub-assignment pair. Nested shadows can also be used to calculate the number of unsatisfying candidate assignments of a 3-SAT problem in depth-first form, recursively. Let $F$ be a 3-SAT problem in $m$ clauses, with $IP$ the family of rank-ordered incompatible sub-assignment pairs of $F$, and $U$ the set of unsatisfying assignments of $F$. For any incompatible sub-assignment pair $ip \in IP$, let $ip.hi$ be the set of unsatisfying candidate assignments in which $ip$ is the highest-ranked incompatible sub-assignment pair, and let $ip.unsh = | ip.hi |$ be the number of unsatisfying candidate assignments in which $ip$ is the highest-ranked incompatible sub-assignment pair. Under these definitions, by the argument of Corollary 6.1, $|U| = \sum_{ip \in IP} ip.unsh$. Thus, with $ip.hi$ defined as the above set,

$$|U| = \sum_{ip \in IP} | ip.hi |. \quad (27)$$

The number of unique unsatisfying candidate assignments of $F$ may also be calculated depth-first, by descending recursively through the shadows of each incompatible sub-assignment pair $ip$, reducing the $8^{m-2}$ unsatisfying candidate assignments in which $ip$ occurs by the number of those unsatisfying candidate assignments in which $ip$ is not highest-ranked. This requires some preliminary definitions. As before, let $IP$ be the set of incompatible sub-assignment pairs of 3-SAT problem $F$. Let $\boldsymbol{IP[\,r\,]}$ denote the set of incompatible sub-assignment pairs of $IP$ that have rank $0 \le r < m(m - 1)/2$. Let $ip \in IP$ be an incompatible sub-assignment pair. The incompatible sub-assignment pairs shadowed by $ip$ are in the set $ip.sh$. Let $\boldsymbol{SC} = (ip_1, \ldots, ip_e)$ be a rank-ordered set of $| SC | = e$ intersecting incompatible sub-assignment pairs over $| SC |$ ranks, where $0 \le | SC | \le m(m - 1)/2$. $SC$ is called an ***e*-shadow**, with $e = | SC |$, if and only if:

1. $| SC | = 0$, (a 0-shadow),
2. $| SC | = 1$, (a 1-shadow), or
3. $| SC | > 1$, and for all pairs $ip, jp \in SC$ such that $ip.rank < jp.rank$, $ip$ shadows $jp$, i.e., $jp \in ip.sh$; (a $| SC |$-shadow). (28)

Let ***IC*** be a set of intersecting sub-assignment pairs. Let $ip = (i.s, j.t)$ be a sub-assignment pair with left and right member sub-assignments *ip.l* and *ip.r*. Under those definitions, $\boldsymbol{S(IC)} = \bigcup_{ip \in IC} ip.l \cup ip.r$, i.e., $S(IC)$ is the union of sub-assignments of the members of *IC*. The number of unsatisfying candidates in which ip is the highest-ranked incompatible sub-assignment pair is denoted $\boldsymbol{\uparrow ip}$, so that $\uparrow ip = |\ ip.hi\ |$. Let *SC* be an *e*-shadow containing sub-assignment pair *ip*. The term, $\boldsymbol{\uparrow ip(\ SC\ )}$, is the number of unsatisfying candidate assignments in which *ip* is highest - ranked under all members of the context *SC*, which may be zero. Disregarding the earlier use of *k* to define *k*-SAT, for some fixed $0 < k \le m(m - 1)/2$, the term ***k-intersection*** denotes a mutually intersecting set of *k* incompatible sub-assignment pairs, such as a lowest-ranked incompatible sub-assignment pair and its $k$ - 1 shadows. A ***k*-shadow** of incompatible sub-assignment pair *ip* is *ip* and the $(k - 1)$ mutually intersecting incompatible sub-assignment pairs that *ip* shadows, (i.e., nested shadows in the set *ip.sh*). The term $|\ \boldsymbol{k\cap}\ |$ is the number of *k*-intersections of *F*.

**Theorem 9.** Let *F* be a 3-SAT problem in *m* clauses, and let *IP* be the set of incompatible sub-assignment pairs of *F*. Let *U* be the set of unsatisfying candidate assignments of *F*. The variable *d* is the recursion depth. Let Ø denote the empty set. Assume that *shadows*(*SC*) is in increasing rank order. For brevity, let *ip.sh* = *ip.children*, (see defs., § 7.) Under those definitions:

$$|\,U\,| = \sum_{0 \le r < m(m-1)/2} \sum_{ip_1 \in IP[r]} \uparrow ip_1(\emptyset), \text{where} \tag{29}$$

$$\uparrow ip_d(SC_d) = 8^{m - |\,S(SC_d \cup ip_d)\,|} - \sum_{ip_{d+1} \in shadows(sc_d \cup\, ip_d)} \uparrow ip_{d+1}(SC_d \cup ip_d), \text{where} \tag{30}$$

(1) $\uparrow ip_{d+1}(SC_d \cup ip_d)$ invokes left-hand term $\uparrow ip_d(SC_d)$ recursively; (31)

$$(2)\quad shadows(SC) = \begin{cases} \emptyset \text{ iff } (\,(SC = \emptyset) \vee ((\bigcap_{ap \in SC} ap.sh) = \emptyset)\,) \\ \qquad otherwise, \bigcap_{ap \in SC} ap.sh \end{cases} \tag{32}$$

(3) $0 < d \le m(m - 1)/2$. (33)

*Proof.* The proof shows the identity of Theorems 9 and 7. Substituting $0 < d \le m(m - 1)/2$ converts Theorem 9 into the equation of Figure 6. Figure 6 is a form of recursive depth-first descent over all *k*-intersections of *F*, each of which is evaluated as the minuend terms of line (1) ... line (*q*), where $0 < k \le m(m - 1)/2$. In line 0 of Figure 6, the Σ Σ invokes | *IP* | instances of $\uparrow ip_1(\emptyset)$, as each $ip_1$ is encountered in the inner Σ. In line 1 of Figure 6, the minuend calculates $\uparrow ip_1(\emptyset) = 8^{m-2}$, the size of a 1-intersection. Then, for each shadow, $ip_2 \in ip_1.sh$, the Σ invokes $\uparrow ip_2(ip_1)$. On line 2, the minuend is $\uparrow ip_2(ip_1) = 8^{m - |S(ip_1,\, ip_2)|}$, the size of a 2-intersection, which is either $\uparrow ip_2(ip_1) = 8^{m-3}$ or $\uparrow ip_2(ip_1) = 8^{m-4}$, depending on the intersection form of $(ip_1, ip_2)$. (Lemma 4.) Then, in the Σ of line 2, for each shadow, $ip_3 \in (ip_1.sh \cap ip_2.sh)$, the term $\uparrow ip_3(ip_1, ip_2)$ is invoked. In line 3 of Figure 6, the minuend, which is $\uparrow ip_3(ip_1, ip_2) = 8^{m - |S(ip_1,\, ip_2,\, ip_3)|}$, is the size of a 3-intersection, i.e., one of: $\{8^{m-3}, 8^{m-4}, 8^{m-5}, 8^{m-6}\}$, (Lemmas 4-5). The Σ of line 3 invokes $\uparrow ip_4(ip_1, ip_2, ip_3)$. In the minuend of line 4 of Figure 6, $\uparrow ip_4(ip_1, ip_2, ip_3)$ evaluates to the size of a 4-intersection, and so on. The chain stops when either: (1) a summation subtrahend iterator finds no joint shadows, or (2) line *q* is reached, because $q = m(m - 1)/2$ is the maximum number of incompatible sub-assignment pairs in a candidate assignment, (by § 2, def. *candidate assignment*).

| ***( Sign )*** | ***Recursion*** | ***= Presumed*** $\uparrow$***ip***$_d$***( ... ) - Σ ( higher ranked shadowed of ip***$_d$**)** | ***d*** | ***Line*** |
|---|---|---|---|---|
| NA | $\|\,U\,\|$ | $= \sum_{0 \le r < \frac{m(m-1)}{2}} \sum_{ip_1 \in IP[r]} \uparrow ip_1(\emptyset)$, where | NA | (0) |
| ( + ) | $\uparrow ip_1(SC_1)$ | $= 8^{m - \|\,S(\,ip_1)\,\|} - \sum_{ip_2 \in shadows(\,ip_1)} \uparrow ip_2(SC_1 \cup ip_1)$, | *d* = 1 | (1) |
| ( - ) | $\uparrow ip_2(SC_2)$ | $= 8^{m - \|\,S(SC_2 \cup\, ip_2)\,\|} - \sum_{ip_3 \in shadows(SC_2 \cup\, ip_2)} \uparrow ip_3(SC_2 \cup ip_2)$, | *d* = 2 | (2) |
| ( + ) | $\uparrow ip_3(SC_3)$ | $= 8^{m - \|\,S(SC_3 \cup\, ip_3)\,\|} - \sum_{ip_4 \in shadows(SC_3 \cup\, ip_3)} \uparrow ip_4(SC_3 \cup ip_3)$, | *d* = 3 | (3) |
| | | … | | |
| | $\uparrow ip_q(SC_q)$ | $= +\,(-1)^{(q+1)} \times 8^{m - \|\,S(SC_q \cup\, ip_q)\,\|}$, where $q = m(m - 1)/2$. | *d* = *q* | (*q*) |

**Figure 6 | Expansion of Theorem 9, by substituting constants for recursive values of *d* = 1...*q***

Thus, the equation of Figure 6 evaluates the size of all $k$-intersections of $F$. The | $IP$ | invocations of $\uparrow ip_1(\varnothing)$ are:

$$| IP | \times 8^{m-2}. \tag{34}$$

By Lemma 5, the invocations of $\uparrow ip_2(ip_1)$ in the depth $d = 1$ summation subtrahends of line 1, evaluated in the minuend of line 2, at depth $d = 2$, total:

$$- [ ( (n_{3(2)} \times 8^{m-3}) + (n_{4(2)} \times 8^{m-4}) ) ]. \tag{35}$$

At $d = 2$, the $\uparrow ip_3(ip_1, ip_2)$ in the summation subtrahend of line 2, evaluated in the line 3, $d = 3$ minuend is:

$$+ [( (n_{3(3)} \times 8^{m-3}) + \ldots + (n_{6(3)} \times 8^{m-6}) )], \tag{36}$$

and so on. That pattern continues for all remaining values of $d$, until, for the summation subtrahends at depth $d = (q - 1)$, evaluated as $q$-shadows at depth $q$, the sum is:

$$+ (-1)^{q+1} \times ((nsmin_{(q)} \times 8^{m - smin_q}) + \ldots + (nsmax_{(q)} \times 8^{m - smax_q})), \text{ where } q = m(m - 1)/2. \tag{37}$$

Assembling lines (34) through (37) and substituting into Figure 6 gives:

$$\begin{aligned} | U | = \; & | IP | \times 8^{m-2} && (d = 1) \\ & - [ ( (n_{3(2)} \times 8^{m-3}) + (n_{4(2)} \times 8^{m-4}) ) ] && (d = 2) \\ & + [( (n_{3(3)} \times 8^{m-3}) + \ldots + (n_{6(3)} \times 8^{m-6}) )] && (d = 3) \\ & \quad \ldots \\ & + (-1)^{q+1} \times ((nsmin_{(q)} \times 8^{\mathrm{m} - smin_q}) + \ldots + (nsmax_{(q)} \times 8^{\mathrm{m} - smax_q})), && (d = q) \\ & \text{where } q = m(m - 1)/2, \end{aligned}$$

proving the identity of Theorem 9 and Theorem 7. (The same result can be achieved by assuming the identity of breadth-first and depth-first search; see [SR98].) □

**Theorem 10.** Let $IP$ be the set of incompatible sub-assignment pair intervals of 3-SAT problem $F$. Let $U$ be the set of unsatisfying candidate assignments of $F$. For $ip \in IP$, let *ip.unshadowed* = *ip.low* be the set of | *ip.unshadowed* | unsatisfying candidate assignments in which $ip$ is an unshadowed shadower, i.e., the set of unsatisfying candidate assignments in which $ip$ is the lowest-ranked incompatible sub-assignment pair.

$$\begin{aligned} | U | &= \textstyle\sum_{ip \in IP} | ip.unshadowed |. \\ | U | &= \textstyle\sum_{ip \in IP} | ip.low |. \end{aligned} \tag{38}$$

*Proof.* Each unsatisfying candidate assignment has exactly one lowest-ranked incompatible sub-assignment pair. (§ 2, def. *rank* and *candidate assignment*.) Incompatible sub-assignment pair $ip$ is only unshadowed in the set of unsatisfying candidate assignments in which $ip$ is the lowest-ranked. (See (14).) Thus, *ip.low* = *ip.unshadowed* and $\forall (ip \neq jp) \in IP \; (ip.low \cap jp.low = \varnothing)$, proving Theorem 10. □

# 10 Discussion, Future Research, Other Abbreviated Findings

The finding that the search space of every 3-SAT problem has a doubly-periodic structure is not entirely surprising. For, Korst proved in Theorem 3.6 of [KJEA94] that every graph is a periodic interval graph. Because sub-assignment pairs are doubly-periodic in matrix $M$, Korst's finding suggests that 3-SAT problems may be described as graphs of graphs. It is beyond the scope of this paper to examine whether such graphs are a subclass of the recursively constructed graphs described in [BRPGTC09], for which there are polynomial-time algorithms to find maximum independent sets.

The system of light and shadows (of §§ 7-9) has physical analogues in source-to-sink or sender-receiver systems in which a (possibly periodic) source or set of source signals can be blocked by a set of (doubly-periodic) interferers, so that the source signals are cancelled at or do not reach the sink (receiver). Such systems have been used for analog and digital noise cancellation. (E.g., see [WBGJMJ75]). There may be analogues in a biological sender-receiver model such as described in [MDSVIM15]. Or, the light and shadows could be cast as a message-passing system, evaluated under a system such as is described in [BAMMZR05].

Space limitations require this paper to abbreviate discussion of properties of the sub-assignment expansion. The following list summarizes those properties, discussed in a longer, companion paper ([PODPOA26]):

1. For every 3-SAT clause pair, either 15, 39, 40, 51, 52, 57, or 58 of its $8^2$ sub-assignment pairs are incompatible, and each incompatible pair rules out its $8^{m-2}$ unsatisfying candidate assignments. The compatibility signature of a 3-SAT problem gives the number of clause pairs in each of those cardinality categories. In common benchmark problems, 91% of clause pairs have 49 compatible sub-assignment pairs, and 9% are more constrained. Suggested uses of compatibility signatures in parallel search and characterizing problem hardness are developed in [PODPOA26].

2. Every *k*-SAT clause has a characteristic finite state automaton of clause compatibility (FSAC), with *m* such instantiated automata per *k*-SAT problem. Every *k*-SAT assignment induces a short path through each instantiated FSAC. Each such path terminates in either a deadlocked (rejecting) or non-deadlocked (accepting) state. Deadlocking combinations can be ruled out without backtracking. Comparing 3-SAT to 2-SAT states indicates how 2-SAT sub-problems may be contained in 3-SAT problems, as analyzed from a different angle in [POD04].

3. Those automata may be used to build a recursively structured triangle of clause pair pigeonholes, in which a satisfying candidate assignment must satisfy certain structural and path properties ([PODPOA26]).

4. Because each incompatible sub-assignment pair is a doubly-periodic signal over $[0, 8^m - 1]$, a 3-SAT problem may also be rendered as superposed periodic signals in which unshadowed (satisfying) candidate assignments are the points at which no signal is present. Such a rendering suggests training machine-learning models to listen to partial sweeps of that superposition, and to predict satisfiability from its periodic texture, much as [SLBLMD19] trained a network to predict satisfiability from a formula's clause graph. Similar superposition patterns arise in 2-SAT (Cor. 5.1, with $w = 4$), for which satisfiability is decidable in linear time ([APT79]). Large sets of labeled 2-SAT problems may therefore be used to train machine-learning models whose predictions may transfer to the presently more difficult 3-SAT problem.Whether either of those techniques is promising is a further question beyond the scope of this paper.

## 11 Exploring the (Wider) Computational Border Between 3-SAT and 2-SAT

With appropriate substitutions (e.g., four sub-assignments per clause instead of eight, and $4^2$ sub-assignment pairs per clause pair instead of $8^2$), the properties in this paper apply to 2-SAT, and to *k*-SAT generally for $k > 0$, with $w = 2^k$.

There are deeper relations between 3-SAT and 2-SAT. For any 3-SAT clause $c_i$, the finite-state automaton of compatibility of $c_i$ contains as its start state a 7-ary set of available sub-assignments. Initially, fixing any single literal of $c_i$ advances the state to one of a pair of subsets: a 4-ary set that is tautological over the remaining two literals of $c_i$, and a 3-ary set that corresponds to a 2-SAT clause, and may thus be viewed as the start state of a 2-SAT automaton ([PODPOA26]). The 3-ary sets of the three pairs of such 4-ary/3-ary sets, which are the start states of 2-SAT sub-automata of $c_i$, correspond to sub-clause populations induced by literal choices, as such populations are defined in [POD04]. It may be inferred that any 3-SAT problem is satisfiable iff at least one candidate assignment satisfies all the 2-SAT sub-automata it induces, confirming a prior finding of [POD04], that every satisfying 3-SAT assignment induces a satisfiable 2-SAT problem of 2-SAT sub-clauses.

3-SAT and 2-SAT are special instances of problems in the sub-assignment domain. Without the *k*-SAT restrictions, for $m \geq 2$, the sub-assignment expansion may be comprised of:

1. a family of *m* (quasi-clause) sets of $w > 0$ sub-assignments each;
2. $0 \leq a \leq w$ anticlause-equivalent (forbidden) sub-assignments for each set;

3. an arbitrary compatibility function, $R$;
4. the other structures such as $m$-ary clause-representative candidate assignments in which a candidate assignment is satisfying iff it contains $m(m - 1)/2$ compatible sub-assignment pairs under $R$;
5. there, a problem instance is satisfiable iff at least one of its candidate assignments is satisfying. (39)

Let that model be called w-SASS. There, the standard 2-SAT and 3-SAT sub-assignment expansions are 4-SASS and 8-SASS, respectively, with each quasi-clause having $a = 1$ anticlause in one of its $w$ positions and compatibility defined by some $R$ that at minimum classifies sub-assignment pairs with forbidden sub-assignments as incompatible. (See [PODPOA26].)

Between 4-SASS and 8-SASS there are at least three similar w-SASS problems—5-SASS, 6-SASS, and 7-SASS—each with one anticlause per quasi-clause and a compatibility function that applies the anticlause rule and defines inconsistency by some other fixed rule. Because 2-SAT is in P ([APT79]) but it is presently unknown whether 3-SAT is in P ([MRPGTR16], [PVNP16], [FL09]), studying these intermediate problems may illuminate the computational boundary between 2-SAT and 3-SAT as a function of $w$, $a$, and $R$, helping to answer the fundamental question:

Is 3-SAT in P?

## REFERENCES

**[APT79]** Aspvall, Bengt; Plass, Michael F.; Tarjan, Robert E. A linear-time algorithm for testing the truth of certain quantified boolean formulas. Information Processing Letters 8 (3):121–123 (1979).

**[BAMMZR05]** Braunstein, A., Mézard, M. and Zecchina, R.: Survey propagation: An algorithm for satisfiability. Random Structures & Algorithms, 27: 201–226 (2005).

**[BRPGTC09]** Borie, R., Parker G., Tovey, C. Solving problems on recursively constructed graphs. ACM Computing Surveys (CSUR), Vol.41:1-51 (Jan. 2009).

**[CLR97]** Cormen, T., *et al*: Introduction to Algorithms. MIT Press, Cambridge (1997):91-98.

**[CMS97]** S.A. Cook and D.G. Mitchell. Finding hard instances of the satisfiability problem: A survey. In Satisfiability Problem: Theory and Applications. DIMACS (Vol. 35):1-12 (1997), American Mathematical Society.

**[DVMPJW2005]** Dahllöf, V., Jonsson, P., Wahlström, M. Counting Models for 2SAT and 3SAT Formulae. Theoretical Computer Science, 332(1-3):265-291 (Feb. 2005).

**[FL09]** Fortnow, Lance. The Status of the P versus NP Problem. Comm. of the ACM, Vol. 52(9):78-86 (Sept. 2009).

**[GP97]** J. Gu, P.W. Purdom, J. Franco, and B.W. Wah. Algorithms for the Satisfiability (SAT) Problem: A Survey. In Satisfiability Problem: Theory and Applications. DIMACS (Vol. 35):19-20 (1997), AMS.

**[HKS03]** Herman, J., Kucera, R., Simsa, J.**:** Counting and Configurations**:** Problems in Combinatorics, Arithmetic, and Geometry (CMS Books in Mathematics), Springer-Verlag (2003): 66-68.

**[HRS10]** D.H. Huson, R. Rupp, and C. Scornavacca. Phylogenetic Networks: Concepts, Algorithms and Applications. Cambridge University Press (2010):8-9.

**[IWA89]** Iwama, K. CNF satisfiability test by counting and polynomial average time. SIAM J. Comput. 18(2):385–391 (1989).

**[KJEA94]** Korst, J., Aarts, E. Periodic assignment and graph colouring. Discrete Applied Mathematics, Vol. 51(3) (6 July 1994): 291-305.

**[KJLNSA96]** Kahn, J., Linial, N., Samorodnitsky, A. Inclusion-Exclusion: Exact and Approximate. Combinatorica, 16(4):465-477 (1996).

**[KUT07]** Konstantin Kutzkov, "New upper bound for the #3-SAT problem," Information Processing Letters 105(1) (2007), 1–5.

**[KUT26]** Konstantin Kutzkov, "Complementary Branching for #SAT: Bounded-Degree Bounds and Faster #3-SAT," in Proceedings of the 35th International Joint Conference on Artificial Intelligence (IJCAI 2026), paper 254. https://www.ijcai.org/proceedings/2026/254.

**[LNNN90]** Linial, N., Nisan, N. Approximate Inclusion-Exclusion. Combinatorica, 10(4):349-365 (1990).

**[MASS97]** Melkman, A.A., Shimony, S.E. A Note on Approximate Inclusion-Exclusion. Discrete Applied Mathematics, 73(1):23-26 (1997).

**[MDSVIM15]** Menendez, D.B., Senthivel, V.R., Isalan, M. Sender–receiver systems and applying information theory for quantitative synthetic biology. Current Opinion in Biotechnology Vol. 31:101-107 (Feb. 2015).

**[MRPGTR16]** Marino, R., Parisi, G., & Ricci-Tersenghi, F. The backtracking survey propagation algorithm for solving random K-SAT problems. Nature Communications 7, (2016).

**[MSKM97]** Muchnick, S. Advanced Compiler Design and Implementation. Morgan Kaufmann (1997).

**[POD04]** D. Powell, "2-Sat Sub-Clauses and the Hypernodal Structure of the 3-Sat Problem," arXiv:cs/0404038 [cs.CC], 2004. https://arxiv.org/abs/cs/0404038.

**[PODPOA26]** D. Powell and A. Powell, "A Common Type Framework for 2-SAT and 3-SAT," Zenodo, 2026. https://doi.org/10.5281/zenodo.23701633.

**[PVNP16]** The P-versus-NP Page. At http://www.win.tue.nl/~gwoegi/P-versus-NP.htm. Last update Sept. 26, 2016.

**[SLBLMD19]** Selsam, D., Lamm, M., Bünz, B., Liang, P., de Moura, L., Dill, D. L. Learning a SAT Solver from Single-Bit Supervision. International Conference on Learning Representations (ICLR) (2019). arXiv:1802.03685.

**[SR98]** Sedgewick, R.: Algorithms in C++, Parts 1-4. Addison-Wesley (1998):255-260.

**[WBGJMJ75]** Widrow, B., Glover, J.R., McCool, J.M., *et al*. Adaptive Noise Cancelling: Principles and Applications. Proc. IEEE, Vol. 63:1692-1716 (Dec. 1975).